\documentclass[aps,prx,showpacs,superscriptaddress,twocolumn,floats,epsfig,longbibliography]{revtex4-2}
\usepackage{amssymb}
\usepackage{amsbsy}
\usepackage{amsmath}
\usepackage{epsfig}
\usepackage{color}
\usepackage{float}
\usepackage{xr-hyper}
\usepackage{bm}
\usepackage[colorlinks,linktocpage,bookmarks=false,citecolor=blue,linkcolor=red,urlcolor=blue]{hyperref}

\newcommand{\beg}{\begin{gather}}
\newcommand{\eeg}{\end{gather}}
\newcommand{\beq}{\begin{equation}}
\newcommand{\eeq}{\end{equation}}
\newcommand{\bea}{\begin{eqnarray}}
\newcommand{\eea}{\end{eqnarray}}

\newcommand{\ket}[1]{\ensuremath{\left|#1\right\rangle}}

\newcommand{\be}{\begin{equation}}
\newcommand{\ee}{\end{equation}}
\def\ba{\begin{aligned}}
\def\ea{\end{aligned}}
\def\bes{\begin{subequations}}
\def\ees{\end{subequations}}
\def\bal{\begin{align}}
\def\eal{\end{align}}

\renewcommand{\hat}[1]{{\widehat #1}}

\usepackage{soul}
\usepackage{ulem} 

\usepackage{hyperref}
\begin{document}

\title{All-Electric Quantum State Transfer via Spin-Orbit Phase Matching}
\author{Madhumita Sarkar}
\thanks{sarkar.madhumita770@gmail.com}
\affiliation{Department of Physics and Astronomy, University College London, Gower Street, WC1E6BT, London.}
\author{Roopayan Ghosh}
\affiliation{Department of Physics and Astronomy, University College London, Gower Street, WC1E6BT, London.}
\affiliation{School of Basic Sciences, Indian Institute of Technology, Bhubaneswar, 752050, India.}
\author{Charles G. Smith}
\affiliation{Hitachi Cambridge Laboratory, J. J. Thomson Avenue, Cambridge CB3 0US, United Kingdom.}
\affiliation{Cavendish Laboratory, Department of Physics, University of Cambridge, Cambridge CB3 0US, United Kingdom.}
\author{Maksym Myronov}
\affiliation{Physics Department, The University of Warwick, Coventry CV4 7AL, UK}
\author{Sougato Bose}
\affiliation{Department of Physics and Astronomy, University College London, Gower Street, WC1E6BT, London.}
\date{\today}

\begin{abstract}

Semiconductor hole-spin qubits offer a promising route to quantum computation due to their weak hyperfine interaction, and strong intrinsic spin–orbit coupling enabling electric control of qubits. Scalable architectures, however, require coherent long-distance quantum state transfer, which is hindered in these systems by spin–orbit-induced anisotropic exchange. Here we show that this limitation can be overcome by using an all-electric control protocol. By tuning the electric field strength, we identify discrete spin–orbit phase-matching conditions that restore near-perfect state transfer, independent of the rotation axis. Complementarily, controlling the electric field direction aligns the spin–orbit axis, suppressing excitation non-conserving processes and enabling robust transfer without fine tuning. Our results establish that electrical control of spin–orbit phases through either magnitude tuning or axis alignment as a practical route for robust quantum information transport in hole-spin quantum dot arrays.
\end{abstract}
\maketitle
\paragraph{\textbf{Introduction--}}
Semiconductor hole-spin qubits have recently emerged as a promising platform for scalable quantum information processing due to their strong spin–orbit coupling (SOC) and the resulting possibility of fast, electric control of qubits and low hyperfine interaction with nuclear spin~\cite{Loss1998, Kloeffel2013, Hendrickx2020, Maurand2016, qgnt-n527}. Rapid experimental progress in gate-defined hole systems has demonstrated high fidelity single and two-qubit operations, electrically tunable g-factors, and ultra-high hole mobilities exceeding $7 \times 10^6$ $\mathrm{cm^{2}V^{-1}s^{-1}}$,~\cite{Geyer2024, Wang2022, Borjans2023, MYRONOV2025314, Myronov2023, Maksym3, Scappucci2021, RevModPhys.95.025003, John2025, 18qubit}. However, extending these platforms toward scalable architectures requires reliable protocols for long-distance quantum state transfer.

Current approaches primarily rely on nearest-neighbor SWAP or iSWAP operations, which necessitate closely spaced qubits and repeated gate operations, thereby increasing susceptibility to crosstalk, noise, and measurement overhead. An alternative strategy is to employ quantum bus architectures, where information is transmitted via coherent by employing an intermediate quantum channel~\cite{ Kurpiers2018, Axline2018, quasi} which can also be a spin chain~\cite{Bose2003, Bose01012007, Bayat2012, PhysRevLett.92.187902}, enabling long-range coupling while mitigating local control constraints..

However, in hole-spin systems strong SOC fundamentally alters the nature of exchange interactions. Instead of an isotropic Heisenberg coupling, the effective interaction becomes anisotropic, incorporating Dzyaloshinskii–Moriya terms and spin rotations that break conventional spin conservation~\cite{Geyer2024}. As a result, the standard intuition for coherent state transfer in spin chains no longer directly applies, and the fidelity of transfer is generally degraded.

Existing schemes to correct SOC-induced spin rotations typically rely on magnetic-field compensation~\cite{Kavokin2001, Trifunovic2012}, which is difficult to scale in hole systems due to strong g-tensor anisotropy and device-specific calibration. Here, we show that purely electrical control provides a scalable alternative: tuning the electric-field strength enables discrete spin–orbit phase matching that restores high-fidelity state transfer independent of the spin–orbit axis orientation, while tuning the field direction stabilizes coherent dynamics by engineering the anisotropic exchange. These mechanisms remain effective under finite magnetic fields relevant for qubit operation, with axis alignment exhibiting greater robustness than discrete phase matching. Together, our results establish an all-electric route toward scalable quantum communication in strongly spin–orbit-coupled platforms.
\begin{figure}[h]
    \centering
    \includegraphics[width=0.98\linewidth]{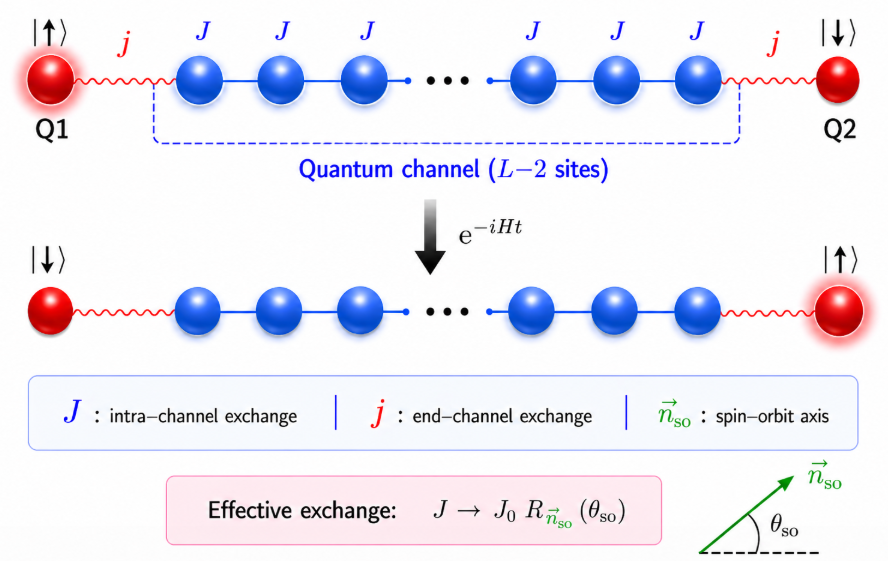}
    \caption{Pictorial representation of our setup for quantum communication. The intermediate qubits shown in blue represents the quantum bus. The other two spins shown in red at the ends (Q1, Q2) represents the qubits of interest.}
    \label{pic1}
\end{figure}

\begin{figure}[ht]
    \centering
    \includegraphics[width=1.\linewidth]{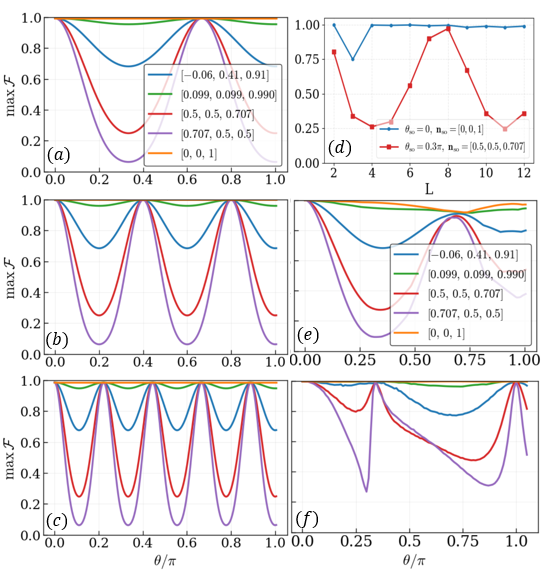}

\caption{
(a)-(c) Maximum state-transfer fidelity (evaluated within a finite time window $t \leq 10\,\mu s$) as a function of the spin-orbit rotation angle $\theta_{\mathrm{so}}$ for different orientations of the spin-orbit axis $\mathbf{n}_{\mathrm{so}}$, shown with different colours as indicated in the leftmost plot. For all system sizes $L=4,6,10$, perfect state transfer is achieved at discrete commensurate values $\theta_{\mathrm{so}} = \frac{2\pi n}{L-1}$, independent of the orientation of $\mathbf{n}_{\mathrm{so}}$. As the system size increases, the number of such optimal points also increases. 
(d) Comparison of the maximum state-transfer fidelity as a function of system size for the isotropic Heisenberg limit ($\theta_{\mathrm{so}}=0$) and for finite spin--orbit coupling with fixed $\theta_{\mathrm{so}}=0.3 \pi$ and $\mathbf{n}_{\mathrm{so}}=[0.5, 0.5, 0.707]$, which render the exchange interaction anisotropic. The anisotropic regime exhibits a pronounced non-monotonic dependence on system size. In particular, the fidelity is strongly enhanced near $L=8$, while lower fidelities are obtained for $L=4$ and $L=11$.
(e) Maximum fidelity plot for pairwise-singlet initialization of the channel. The behavior remains qualitatively similar to the ground-state case, with near-perfect transfer recovered at the commensurate point for $L=4$. 
(f) Effect on fidelity of a finite magnetic field ($\mathbf{B}\sim 50\,\mathrm{MHz}$) used to define the spin quantization axis. High-fidelity transfer persists for all orientations of $\mathbf{n}_{\mathrm{so}}$, although the optimal value of $\theta_{\mathrm{so}}$ is shifted, indicating a field-dependent modification of the phase-matching condition.
For all plots $J_0=160\,\mathrm{MHz}$ and $j_0=20\,\mathrm{MHz}$.
}
    \label{fig:fig6a}
\end{figure}

\paragraph{\textbf{Setup--}} Hole spins confined in semiconductor quantum dots occupy predominantly $p$-type orbitals and therefore experience strong intrinsic spin-orbit coupling (SOC). As a result, tunnelling between neighbouring sites becomes spin dependent. We model the system using an extended Fermi-Hubbard Hamiltonian, which, unlike the standard model, has an extra spin dependent tunneling term,
\begin{eqnarray}
H_{FH} =&
\sum_{\langle ij\rangle}\sum_{ss'}
c^\dagger_{is}\, T^{ss'}_{ij}\, c_{js'}
+
U\sum_i n_{i\uparrow}n_{i\downarrow}\\ \nonumber
+& \frac{1}{2}\,\mu_B \sum_i \sum_{s s'} 
c^\dagger_{i s} (\mathbf{B} \cdot \hat{g}_i \cdot \boldsymbol{\sigma})_{s s'} c_{i s'},
\end{eqnarray}
where $c^\dagger_{is}$ ($c_{is}$) creates (annihilates) a hole with spin $s$ on site $i$, and $n_{is}=c^\dagger_{is}c_{is}$ is the spin-resolved number operator. The vector $\mathbf{B}$ denotes the applied magnetic field, and $\mu_B$ is used to denote Bohr magneton, which we set to unity throughout. The tensors $\hat{g}_i$ represent the (generally anisotropic) Land\'e $g$-factors of the individual quantum dots. The matrix elements $T^{ss'}_{ij}$ describe spin-dependent tunnelling between neighbouring sites.

In the presence of SOC the tunnelling matrix can be written as
\begin{eqnarray}
T^{ss'}_{ij}
=
t_{ij}\delta_{ss'}
+
i\sum_{\alpha=x,y,z}\lambda^\alpha_{ij}(\sigma^\alpha)_{ss'},
\end{eqnarray}
where $t_{ij}$ is the spin-conserving hopping amplitude, $\sigma^\alpha$ are the Pauli matrices, and $\boldsymbol{\lambda}_{ij}=(\lambda^x_{ij},\lambda^y_{ij},\lambda^z_{ij})$ encodes the strength and direction of the spin--orbit interaction.

In the strongly interacting regime \( U \gg |t_{ij}|, |\boldsymbol{\lambda}_{ij}|\), double occupation is suppressed and the low-energy sector is restricted to one hole per site. Performing a Schrieffer--Wolff transformation yields an effective spin Hamiltonian~\cite{Geyer2024},
\begin{equation}
H = \sum_{i,j,\langle i,j \rangle} \left[
\frac{1}{2}\mu_B \, \mathbf{B} \cdot \hat{g}_i \cdot \boldsymbol{\sigma}_i 
+ \frac{1}{4} \, \boldsymbol{\sigma}_i \cdot \hat{J} \cdot \boldsymbol{\sigma}_j
\right].
\label{eq:aniso}
\end{equation}

Here, \( \hat{J} = \mathcal{J}_0 \, R(\theta_{so}, \mathbf{n}_{so}) \) is the anisotropic exchange tensor, determined by a rotation matrix \( R(\theta_{so}, \mathbf{n}_{so}) \) describing a rotation by an angle \( \theta_{so} \) about the axis \( \mathbf{n}_{so} \). The vector \( \mathbf{n}_{so} \) specifies the orientation of the spin--orbit axis, while \( \theta_{so} = 2d/\lambda_{\rm so} \), where \( d \) is the inter-dot distance and \( \lambda_{\rm so} \) is the spin--orbit length. In the absence of spin-dependent hopping (\( \boldsymbol{\lambda}_{ij} = 0 \)), the exchange tensor reduces to the isotropic Heisenberg interaction. We consider a weak magnetic field $\mathbf{B}$, sufficient only to define the spin quantisation axis.



Our setup is shown in Fig.~\ref{pic1}. The sender and receiver qubits, located at sites $1$ and $N$, are weakly coupled (with strength $j$) to an intermediate chain of spins that forms the quantum channel, where the intra-channel exchange is stronger ($J$). This geometry enables spatial separation of the computational qubits while allowing coherent quantum information transfer through the spin chain.

The performance of quantum state transfer between the end qubits is quantified by the fidelity. We initialize the sender qubit in $|\uparrow\rangle$ and the receiver qubit in $|\downarrow\rangle$, while the channel is prepared either in its many-body ground state via adiabatic evolution~\cite{adiabatic1,adiabatic2} or via non-adiabatic techniques ~\cite{Long2025}. The system then evolves under $H$, and the receiver state is obtained from the reduced density matrix $\rho_N(t)$. The transfer fidelity is defined as
$F\psi(t) = \langle \psi | \rho_N(t) | \psi \rangle$ ~\cite{Bose2003}
which measures the probability of successful transfer of the spin state.

 


\paragraph{\textbf{Electric-Field Control of State Transfer}-}
Intrinsic spin-orbit interaction modifies the exchange coupling between neighboring spins by generating anisotropic terms beyond the Heisenberg limit [Eq.~(\ref{eq:aniso})]. In particular, terms such as $\sigma^x\sigma^z$ and $\sigma^y\sigma^z$ induce single-spin-flip processes that mix different excitation sectors, creating unwanted excitations and suppressing coherent state transfer. Crucially, these interactions are not arbitrary but arise from a structured rotation of the underlying Heisenberg exchange, characterized by a spin-orbit phase $\theta_{\mathrm{so}}$ and rotation axis $\mathbf{n}_{\mathrm{so}}$. The spin-orbit phase is controlled by the electric-field strength, while the rotation axis is set by the field direction. In what follows, we show how this structure can be exploited to recover high-fidelity quantum state transfer.

\textit{(a) Controlling the strength of electric field:} 
We first focus on controlling the electric-field strength, which tunes the spin-orbit phase $\theta_{\mathrm{so}}$. The rotation angle governing the effective exchange interaction is given by $\theta_{\mathrm{so}} = 2d/\lambda_{\mathrm{so}}$, where $d$ is the inter-dot spacing and $\lambda_{\mathrm{so}}$ is the spin-orbit length. In Ge hole quantum dots, the dominant electrically tunable spin-orbit interaction typically arises from direct Rashba coupling mediated by heavy-hole and light-hole mixing~\cite{rashba}. This generates an effective spin-orbit coupling strength $\alpha_{\mathrm{so}}$ that increases approximately linearly with the applied electric field, $\alpha_{\mathrm{so}} \propto |E|$. Consequently, the spin-orbit length decreases as $\lambda_{\mathrm{so}} \sim \hbar^2/(m^*\alpha_{\mathrm{so}})$, where $m^*$ is the effective mass, leading to $\theta_{\mathrm{so}} \propto |E|$. This enables continuous electrical control of the exchange-induced spin rotation using gate voltages without modifying the device geometry.

To quantify the effect of electric-field tuning, we vary $\theta_{\mathrm{so}}$ and plot the resulting state-transfer fidelity in Fig.~\ref{fig:fig6a}(a)--(c). Remarkably, we find discrete values of $\theta_{\mathrm{so}}$ at which high-fidelity transfer is exactly restored, with the locations of these peaks depending on the system size. Specifically, perfect state transfer occurs when
$\theta_{\mathrm{so}} = \frac{2\pi n}{L-1}, \qquad n \in \mathbb{Z},
\label{phase}$
where $L$ denotes the total number of spins, including the sender, receiver, and intermediate channel qubits. 

This condition emerges because each exchange link introduces a spin rotation by $\theta_{\mathrm{so}}$, such that a spin excitation traversing the chain accumulates a total rotation of $(L-1)\theta_{\mathrm{so}}$. When this net rotation equals an integer multiple of $2\pi$, $(L-1)\theta_{\mathrm{so}} = 2\pi n$, 
the sender and receiver qubits are effectively aligned in the same spin frame, allowing the transferred state to arrive without an additional spin mismatch. For experimentally relevant Ge hole qubits, where $\lambda_{\mathrm{so}}$ can be tuned over a wide range through gate voltages, these phase-matching points are readily accessible without modifying the device geometry.

Remarkably, this condition is independent of the orientation of the spin–orbit axis $\mathbf{n}_{\mathrm{so}}$. This implies that tuning the electric field strength alone acts as a \textit{universal control knob}, capable of recovering perfect state transfer even in the presence of arbitrary anisotropy induced by spin–orbit interaction. In contrast to protocols that require precise engineering of multiple Hamiltonian parameters, here a single experimentally accessible parameter suffices.

As shown in Fig.~\ref{fig:fig6a}(a)-(c), increasing the system size $L$ increases the number of commensurate points, providing additional operating points for high-fidelity transfer. However, these fidelity peaks simultaneously become narrower, requiring more precise tuning of the electric field. This reflects a key tradeoff of the phase-matching mechanism: larger systems offer more accessible commensurate points, but demand tighter control over the spin-orbit phase. 

This trend is further reflected in Fig.~\ref{fig:fig6a}(d), where the transfer fidelity is compared across system sizes. In the Heisenberg limit, the fidelity remains consistently high and is only weakly dependent on $L$. In contrast, generic spin-orbit coupling suppresses the fidelity and introduces a pronounced non-monotonic dependence on system size due to the accumulated spin rotations across the chain. Interestingly, specific system sizes such as $L=8$ recover fidelities close to the Heisenberg limit because the corresponding accumulated phase lies near an effective commensurate point for the chosen parameters. 
\begin{figure}[h]
    \centering
    \includegraphics[width=01.\linewidth]{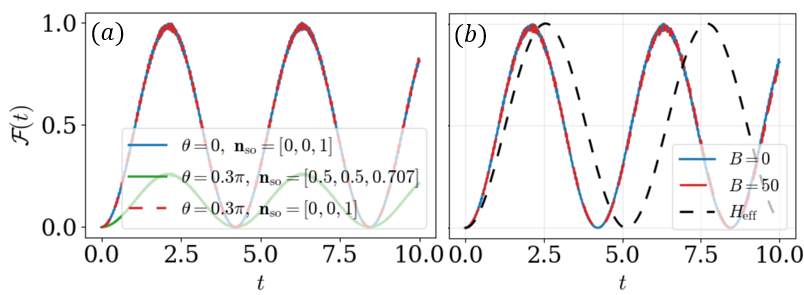}
   
\caption{
State-transfer fidelity $\mathcal{F}(t)$ as a function of time for different spin--orbit configurations. 
(a) Comparison of dynamics for different spin--orbit axes. In the isotropic Heisenberg limit ($\theta_{\mathrm{so}}=0$), the fidelity exhibits clean periodic oscillations approaching unity (blue), corresponding to ideal coherent state transfer. For a generic tilted spin--orbit axis $\mathbf{n}_{\mathrm{so}}=[0.5,\,0.5,\,0.707]$ with $\theta_{\mathrm{so}}=0.3\pi$, anisotropic exchange components suppress the fidelity due to mixing of excitation sectors (green). In contrast, aligning the spin--orbit axis along the $z$-direction, $\mathbf{n}_{\mathrm{so}}=[0,\,0,\,1]$, restores high-fidelity oscillations despite finite spin-orbit coupling (red), demonstrating that electric-field control of the spin-orbit axis enables recovery of ideal state transfer. 
(b) Robustness of axis alignment in the presence of a finite magnetic field. The fidelity dynamics for $B=50$ (red) remain essentially unchanged compared to the zero-field case (blue), confirming that the axis-alignment protocol is insensitive to magnetic-field strength. The dashed curve shows the analytical result obtained from the effective Hamiltonian, which is in agreement with the numerical simulation.
}
    \label{fig4}
\end{figure}

Having established the ideal phase-matching condition, we now examine its robustness under experimentally relevant constraints. Adiabatic preparation of the many-body ground state can consume valuable coherence time, motivating a faster initialization protocol where the channel is prepared in pairwise singlet states. As shown in Fig.~\ref{fig:fig6a}(e), the fidelity profile for $L=4$ remains qualitatively similar to the ground-state case, with near-perfect transfer recovered at the commensurate point $\theta_{\mathrm{so}}=2\pi/3$. This demonstrates that the phase-matching mechanism remains effective even under simplified state preparation protocols compatible with current quantum-dot architectures.

We further account for the finite magnetic field required to define the spin quantization axis. As shown in Fig.~\ref{fig:fig6a}(f), high-fidelity transfer persists and remains largely independent of the orientation of $\mathbf{n}_{\mathrm{so}}$, although the optimal value of $\theta_{\mathrm{so}}$ shifts with the magnetic-field strength. The modified phase-matching condition can be computed for a given field strength, as shown in Appendix~\ref{app1}, and is connected to the emergence of a transport doublet which we discuss in Appendix.~\ref{app:twoqubit_channel}.

For experimentally relevant Ge hole-spin quantum dots, spin–orbit lengths in the range $\lambda_{so} \sim 20–200nm$ have been reported, depending on confinement geometry and vertical electric field strength, while typical inter-dot spacings are $d \sim 50\text{–}150 \,\mathrm{nm}$. This enables electrically accessible spin–orbit rotation angles spanning $\theta_{so} \sim 0.1\pi\text{–}2\pi$ using gate-voltage variations in the experimentally realistic range of a few to several tens of millivolts, making the phase-matching conditions identified here directly compatible with current Ge quantum-dot architectures~\cite{Hendrickx2020}.

But at the same time, the sensitivity of the finely-tuned operating points motivates a complementary strategy based on controlling the spin-orbit axis, which enables robust transfer over a broader parameter regime.

\textit{(b) Controlling the direction of the Electric Field:}
We now turn to a complementary form of phase control, where the direction of the applied electric field determines the orientation of the spin--orbit axis $\mathbf{n}_{\mathrm{so}}$. As discussed earlier, the spin-orbit axis is set by
\begin{equation}
\mathbf{n}_{\mathrm{so}} \propto \mathbf{k} \times \mathbf{E},
\end{equation}
where $\mathbf{k}$ denotes the momentum direction of the quantum dot array and $\mathbf{E}$ is the applied electric field. This provides a direct and experimentally accessible route to control the structure of the effective exchange interaction through gate-defined electric fields.

The orientation of $\mathbf{n}_{\mathrm{so}}$ determines which anisotropic exchange terms are generated and therefore strongly influences the transfer dynamics. For generic orientations, terms such as $\sigma^x\sigma^z$ and $\sigma^y\sigma^z$ induce single spin-flip processes that create excitation leakage and suppress transfer fidelity. In contrast, when $\mathbf{n}_{\mathrm{so}}$ is aligned along the $z$ direction, these terms are eliminated and the dynamics are governed by a more constrained form of anisotropic exchange. As we show below, this suppresses the dominant leakage channels and enables robust high-fidelity quantum state transfer over a broad parameter regime without requiring fine tuning of $\theta_{\mathrm{so}}$.

\begin{figure}[ht]
    \centering
    \includegraphics[width=1.\linewidth]{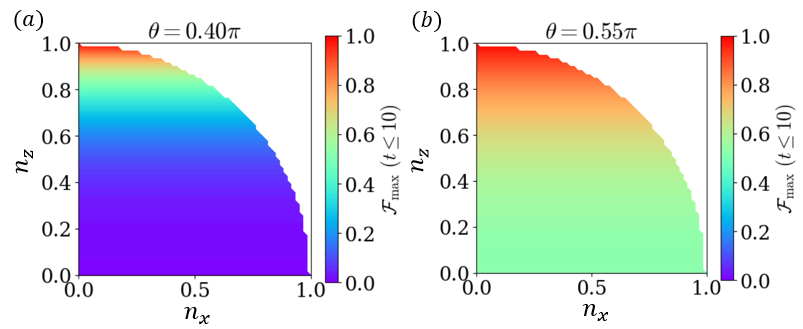}
    \caption{
Maximum state-transfer fidelity within a chosen coherence window ($t \leq 10$) as a function of the spin-orbit axis components $n_x$ and $n_z$ (with $|\mathbf{n}_{\mathrm{so}}|=1$) for two representative rotation angles: (a) $\theta = 0.4\pi$, (b) $\theta = 0.55\pi$. These results are for $L=4$. In both cases, the fidelity is maximized when the spin-orbit axis aligns close to the $z$-direction.
}
    \label{fig7a}
\end{figure}
This behavior is illustrated in Fig.~\ref{fig4}(a), where aligning $\mathbf{n}_{\mathrm{so}}$ along the $z$ direction leads to high-fidelity transfer over a broad range of $\theta_{\mathrm{so}}$, in contrast to the sharply tuned commensurate points discussed earlier. This configuration leaves a reduced anisotropic exchange structure dominated by the diagonal terms and the $J_{xy},J_{yx}$ couplings.

The origin of this robustness can be understood from a two-spin calculation retaining the diagonal components $J_{xx},J_{yy},J_{zz}$ together with the $J_{xy}$ and $J_{yx}$ terms. For the state-transfer protocol considered here, the sender and receiver are initialized in the single-spin-flip sector relative to the antiferromagnetic channel, making the subspace $\{|\uparrow\downarrow\rangle,|\downarrow\uparrow\rangle\}$ the relevant manifold for the transfer dynamics. In this reduced model, the dominant leakage channels are absent and the evolution remains confined within this subspace. Starting from $|\uparrow\downarrow\rangle$, the evolved state is given by
\begin{eqnarray}
|\psi_f(t)\rangle
&=&
e^{iJ_{zz}t}
\Bigg[
\cos(\Omega t)|\uparrow\downarrow\rangle \nonumber\\
&+&
\frac{-i(J_{xx}+J_{yy})-(J_{xy}-J_{yx})}{\Omega}
\sin(\Omega t)
|\downarrow\uparrow\rangle
\Bigg],
\end{eqnarray}
where
\begin{equation}
\Omega=\sqrt{(J_{xy}-J_{yx})^2+(J_{xx}+J_{yy})^2}.
\end{equation}

This describes coherent oscillations between the two spin configurations without the dominant leakage processes present for generic orientations of $\mathbf{n}_{\mathrm{so}}$. In larger chains, while additional many-body processes remain present, the absence of single-spin-flip terms continues to strongly suppress leakage and leads to robust high-fidelity transfer across a broad parameter regime. The oscillation frequency can be estimated using Van-Vleck perturbation theory (Appendix~\ref{app3}), and the resulting analytical prediction agrees well with numerics, as shown by the black dashed line in Fig.~\ref{fig4}(b).

A key advantage of this protocol is its robustness to finite magnetic fields. Unlike the discrete phase-matching mechanism, where the optimal value of $\theta_{\mathrm{so}}$ shifts with magnetic-field strength, the axis-alignment protocol remains largely unaffected, as shown in Figs.~\ref{fig4}(b) and \ref{fig7a}. In particular, the magnetic field does not modify the dynamics when $\mathbf{B} \parallel \mathbf{n}_{\mathrm{so}}$. Since the magnetic field is applied along the $z$-direction to define the spin quantisation axis, and optimal state transfer is achieved when $\mathbf{n}_{\mathrm{so}}$ is aligned along $z$, the protocol remains insensitive to the presence of the magnetic field. This robustness is especially relevant for hole-spin qubits, where strong $g$-factor anisotropy can lead to spatially varying effective magnetic fields, making repeated recalibration of phase-matching conditions experimentally challenging. More details are available in Appendix~\ref{app1}. 
For a quantum dot array oriented along $\mathbf{k}\parallel\hat{x}$, the optimal regime $\mathbf{n}_{\mathrm{so}}\parallel\hat{z}$ can be realized simply by applying the electric field along $\hat{y}$, providing a practical route to robust all-electric quantum state transfer.
We have also verified that the protocol remains robust against charge noise, the dominant low-temperature decoherence mechanism in gate-defined semiconductor spin qubits. Since electrical fluctuations primarily renormalize the exchange couplings, their leading effect is to shift the oscillation frequency and optimal transfer time rather than destroy the underlying transfer mechanism. Full numerical simulations for correlated and uncorrelated noise realizations show that high-fidelity transfer remains achievable over experimentally relevant noise strengths, which we discuss in Appendix~\ref{app4}.

\paragraph{\textbf{Discussion and Outlook:}}
We have shown that intrinsic spin-orbit interaction, often regarded as a major obstacle to coherent transport in hole-spin systems, can instead be transformed into a resource through electrical control. By tuning the electric-field strength, we identify discrete phase-matching conditions that recover perfect state transfer, while controlling the field direction enables a broader and more robust operating regime by suppressing the dominant leakage channels.

These results suggest a natural architecture for scalable hole-spin processors based on two-dimensional arrays of qubits connected by horizontal and vertical spin buses. In such a square network, global in-plane electric fields with tunable $E_x$ and $E_y$ components can be used to route quantum information along both directions. For transport along a given link, only the electric-field component transverse to the propagation direction contributes to $\mathbf{n}_{\mathrm{so}} \propto \mathbf{k}\times\mathbf{E}$, allowing the spin-orbit axis to remain aligned along $\hat{z}$ across the network. The electric-field strength can then be tuned independently to operate near commensurate phase-matching points, enabling robustness and optimal fidelity to be achieved simultaneously through both control protocols.

More broadly, controlling spin--orbit-induced rotations during transport preserves spin coherence while introducing an additional electrical degree of freedom for engineering quantum operations. This opens new opportunities for entanglement distribution, quantum routing, and scalable all-electric quantum computing architectures. Our work establishes spin-orbit engineering as a powerful paradigm for quantum information processing in semiconductor platforms.

\section*{Acknowledgements}
All authors
acknowledge EPSRC-SFI funded project EP/X039889/1
(GeQuantumBus).

\appendix
\section{Effect of Magnetic Field}
\label{app1}

To assess the role of an external magnetic field, we consider the Hamiltonian
\begin{eqnarray}
H = \sum_{i, j, \langle i,j \rangle} \left[
\frac{1}{2}\mu_B \, \mathbf{B} \cdot \hat{g}_i \cdot \boldsymbol{\sigma}_i 
+ \frac{1}{4} \, \boldsymbol{\sigma}_i \cdot \hat{J} \cdot \boldsymbol{\sigma}_j
\right],
\end{eqnarray}
where the first term describes the Zeeman interaction in the presence of an anisotropic $g$-tensor and the second term represents the exchange interaction. We take $\mu_B=1$ and use experimentally motivated $g$-tensors, with $\hat{g}_1$ for the boundary qubits and $\hat{g}_2$ for the channel.

\begin{figure*}[t]
    \centering
    \includegraphics[width=0.7\linewidth]{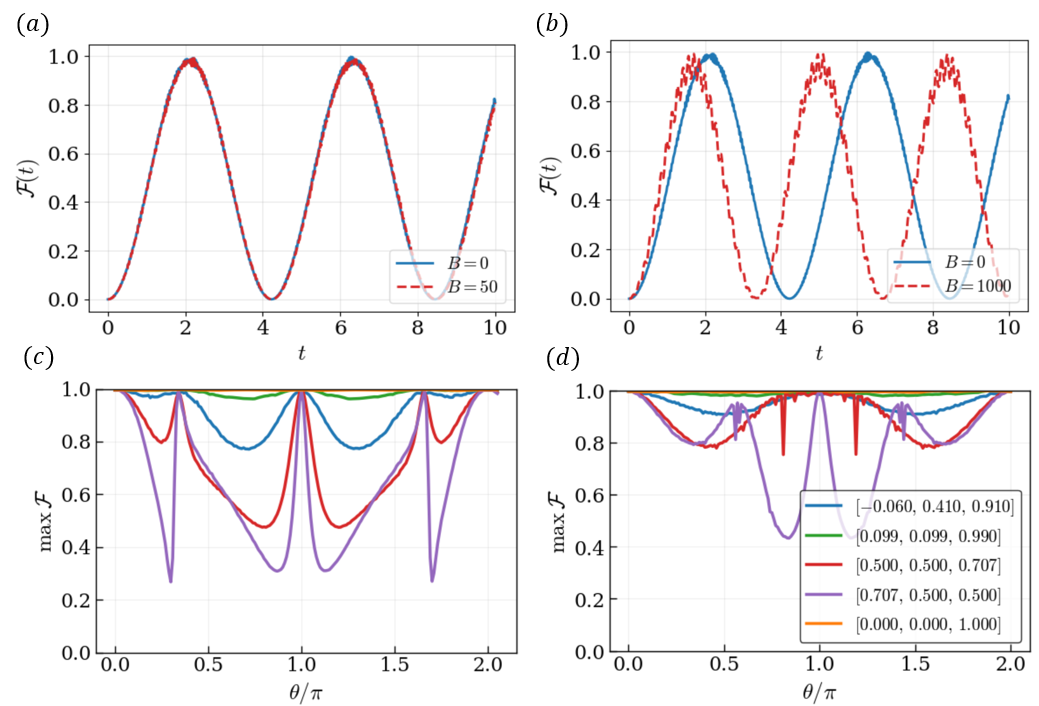}
     \includegraphics[width=0.7\linewidth]{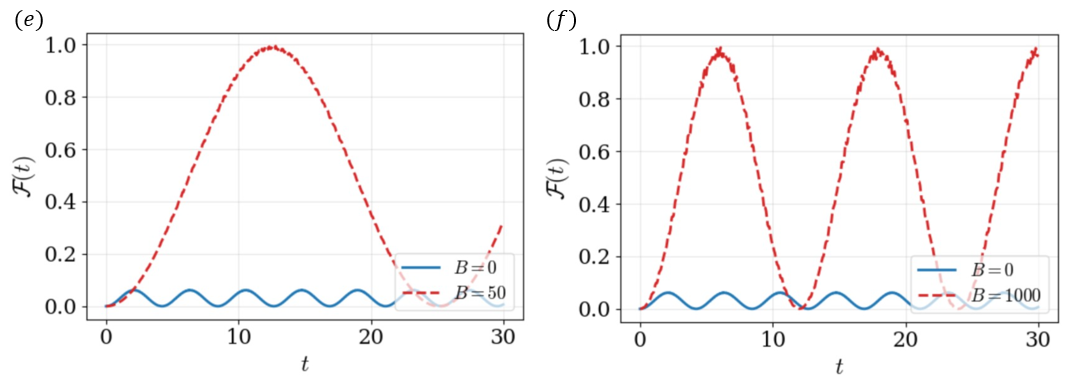}
     
\caption{
(a,b) Time evolution of the fidelity $\mathcal{F}(t)$ for moderate ($B=50\,\mathrm{MHz}$) and large ($B=1000\,\mathrm{MHz}$) magnetic fields in the optimal axis-alignment regime. For experimentally relevant fields ($B=50\,\mathrm{MHz}$), the dynamics remain essentially identical to the zero-field case, while larger fields introduce visible deviations only at longer times. In both cases, however, the maximum fidelity remains unchanged, demonstrating that the axis-alignment protocol is robust against variations in magnetic-field strength. (c,d) Maximum fidelity within a fixed time window ($t<10$) as a function of $\theta_{\mathrm{so}}$ for different orientations of $\mathbf{n}_{\mathrm{so}}$, shown for (c) $B=50\,\mathrm{MHz}$ and (d) $B=1000\,\mathrm{MHz}$. While the axis-alignment regime remains robust in the presence of a magnetic field, the discrete phase-matching points are quantitatively shifted, indicating a field-dependent modification of the optimal $\theta_{\mathrm{so}}$. (e,f) Fidelity dynamics at $\theta_{\mathrm{so}}=\pi$ for $\mathbf{n}_{\mathrm{so}}=[0.5,\,0.5,\,0.707]$. Perfect state transfer is recovered independent of both the magnetic-field strength and the orientation of the spin--orbit axis (here it is $\mathbf{n}_{so}=[0.707, 0.5, 0.5]$), revealing an additional magnetic-field-induced phase-matching condition at $\theta_{\mathrm{so}}=\pi$.
}

    \label{figmag_new}
\end{figure*}

Figures~\ref{figmag_new}(a,b) show the time evolution of the fidelity in the presence of a magnetic field. For experimentally relevant values $B \sim 50$ MHz [Fig.~\ref{figmag_new}(a)], which are sufficient to define the spin quantisation axis, the dynamics are essentially indistinguishable from the zero-field case within the time window of interest. This justifies the approximation used in the main text, where only the exchange interaction is retained. For larger fields ($B \sim 1000$ MHz), the dynamics exhibit quantitative deviations, primarily in the form of phase shifts, but the overall behavior and main conclusions remain unchanged. In particular, the axis-alignment mechanism continues to yield high-fidelity transfer, demonstrating its robustness against the presence of a magnetic field.

We further analyze the dependence of the maximum fidelity on the spin--orbit parameter $\theta_{\mathrm{so}}$ in Figs.~\ref{figmag_new}(c,d). For experimentally relevant magnetic fields [Fig.~\ref{figmag_new}(c)], the phase-matching condition is modified only quantitatively: the optimal values of $\theta_{\mathrm{so}}$ exhibit a slight shift, while the overall structure and achievable fidelity remain unchanged. 

For completeness, we also consider larger magnetic fields ($B \sim 1000$ MHz) in Fig.~\ref{figmag_new}(d), where the deviations become more pronounced. In this regime, precise knowledge of the phase-matching points is required for optimal performance. Nevertheless, the qualitative behavior persists, and the phase-matching protocol continues to yield high-fidelity, and in particular near-perfect, state transfer.

Taken together, these results demonstrate that our protocol is robust to realistic magnetic fields. In particular, a small magnetic field sufficient to define the spin quantisation axis ($B \sim 50$ MHz) does not affect the dynamics, thereby justifying the approximation used in the main text, where only the exchange interaction is considered.

\section{Two-Qubit Channel Hamiltonian in the Presence of Magnetic Field}
\label{app:twoqubit_channel}

To understand the role of magnetic field on the channel spins, we isolate the central two-spin subsystem (sites $2,3$), described by

\begin{equation}
H_{23}
=
J\,\mathbf{S}_2^{T}R(\hat{n},\theta_{\rm so})\mathbf{S}_3
+
(g_2\mathbf B)\cdot(\mathbf S_2+\mathbf S_3),
\label{eq:H23_general}
\end{equation}

where $J$ is the exchange coupling, $\theta_{\rm so}$ is the spin-orbit angle, and

\begin{equation}
R(\hat n,\theta_{\rm so})
=
\cos\theta_{\rm so}I
+
(1-\cos\theta_{\rm so})\hat n\hat n^T
+
\sin\theta_{\rm so}N
\end{equation}

is the spin-orbit rotation matrix with

\begin{equation}
N_{ab}=\epsilon_{abc}n_c.
\end{equation}

We consider a magnetic field applied along the laboratory $z$ direction,

\begin{equation}
\mathbf B=(0,0,B),
\end{equation}

such that the effective field on the channel spins becomes

\begin{equation}
\mathbf h=g_2\mathbf B
=
B
\begin{pmatrix}
g_{xz}\\
g_{yz}\\
g_{zz}
\end{pmatrix},
\end{equation}

or equivalently

\begin{equation}
h_x=Bg_{xz},
\qquad
h_y=Bg_{yz},
\qquad
h_z=Bg_{zz}.
\end{equation}

In the computational basis

\begin{equation}
\{
|\uparrow\uparrow\rangle,
|\uparrow\downarrow\rangle,
|\downarrow\uparrow\rangle,
|\downarrow\downarrow\rangle
\},
\end{equation}

the exact Hamiltonian takes the form

\begin{widetext}
\begin{equation}
\begin{pmatrix}
\frac{J}{4}R_{zz}+h_z
&
\frac{J}{4}(R_{zx}-iR_{zy})+\frac{h_x-ih_y}{2}
&
\frac{J}{4}(R_{xz}-iR_{yz})+\frac{h_x-ih_y}{2}
&
\frac{J}{4}(R_{xx}-R_{yy}-i(R_{xy}+R_{yx}))
\\[1.2em]

\frac{J}{4}(R_{zx}+iR_{zy})+\frac{h_x+ih_y}{2}
&
-\frac{J}{4}R_{zz}
&
\frac{J}{4}(R_{xx}+R_{yy}+i(R_{xy}-R_{yx}))
&
\frac{J}{4}(R_{xz}-iR_{yz})+\frac{h_x-ih_y}{2}
\\[1.2em]

\frac{J}{4}(R_{xz}+iR_{yz})+\frac{h_x+ih_y}{2}
&
\frac{J}{4}(R_{xx}+R_{yy}-i(R_{xy}-R_{yx}))
&
-\frac{J}{4}R_{zz}
&
\frac{J}{4}(R_{zx}-iR_{zy})+\frac{h_x-ih_y}{2}
\\[1.2em]

\frac{J}{4}(R_{xx}-R_{yy}+i(R_{xy}+R_{yx}))
&
\frac{J}{4}(R_{xz}+iR_{yz})+\frac{h_x+ih_y}{2}
&
\frac{J}{4}(R_{zx}+iR_{zy})+\frac{h_x+ih_y}{2}
&
\frac{J}{4}R_{zz}-h_z
\end{pmatrix}.
\label{eq:full_channel_matrix}
\end{equation}
\end{widetext}

The off-diagonal terms encode spin-orbit-induced spin flips and anisotropic mixing, while $h_z$ energetically favors spin polarization.

\subsection{The channel ground state}

\paragraph{Zero magnetic field.}

For $B=0$, the Zeeman contribution vanishes,

\begin{equation}
h_x=h_y=h_z=0,
\end{equation}

and the channel dynamics is governed entirely by the exchange interaction. In this limit, Eq.~(\ref{eq:full_channel_matrix}) reduces to a purely exchange-driven four-level problem in the basis

\begin{equation}
\{
|\uparrow\uparrow\rangle,
|\uparrow\downarrow\rangle,
|\downarrow\uparrow\rangle,
|\downarrow\downarrow\rangle
\}.
\end{equation}

The dominant antiferromagnetic exchange processes act within the subspace

\begin{equation}
\{
|\uparrow\downarrow\rangle,
|\downarrow\uparrow\rangle
\},
\end{equation}

for which the projected Hamiltonian is

\begin{widetext}
\begin{equation}
H_{\rm AFM}=
\begin{pmatrix}
-\frac{J}{4}R_{zz}
&
\frac{J}{4}\left(R_{xx}+R_{yy}+i(R_{xy}-R_{yx})\right)
\\
\frac{J}{4}\left(R_{xx}+R_{yy}-i(R_{xy}-R_{yx})\right)
&
-\frac{J}{4}R_{zz}
\end{pmatrix}.
\label{eq:AFM_block}
\end{equation}
\end{widetext}

The corresponding eigenstates are

\begin{equation}
|\chi_{\pm}\rangle=
\frac{
|\uparrow\downarrow\rangle
\pm
e^{i\phi}
|\downarrow\uparrow\rangle
}{\sqrt{2}},
\end{equation}

where

\begin{equation}
e^{i\phi}
=
\frac{
R_{xx}+R_{yy}+i(R_{xy}-R_{yx})
}{
\left|
R_{xx}+R_{yy}+i(R_{xy}-R_{yx})
\right|
}.
\end{equation}

These states reduce to the usual singlet-triplet combinations in the absence of spin-orbit coupling. However, for finite $\theta_{\rm so}$ the remaining off-diagonal terms in Eq.~(\ref{eq:full_channel_matrix}) generate additional single-spin and double-spin flip processes that couple the antiferromagnetic sector to the ferromagnetic states $|\uparrow\uparrow\rangle$ and $|\downarrow\downarrow\rangle$. Consequently, the exact channel ground state at $B=0$ is generally a superposition of all four computational basis states, with the relative weights determined by both the spin-orbit angle and the rotation axis.

Importantly, although spin-orbit coupling modifies the structure of the eigenvectors, it does not introduce any linear spin-polarization term. The full Hamiltonian therefore remains invariant under time-reversal symmetry,

\begin{equation}
\mathcal{T}H\mathcal{T}^{-1}=H,
\end{equation}

which prevents the strong energetic separation of the low-energy manifold observed at finite magnetic field.

\paragraph{Finite magnetic field.}

For finite field,

\begin{equation}
E_{\uparrow\uparrow}
=
\frac{J}{4}R_{zz}+h_z,
\end{equation}

\begin{equation}
E_{\downarrow\downarrow}
=
\frac{J}{4}R_{zz}-h_z.
\end{equation}

Increasing $B$ lowers the energy of $|\downarrow\downarrow\rangle$, while the antiferromagnetic sector remains centered near

\begin{equation}
E_{\rm AFM}\sim -\frac{J}{4}R_{zz}.
\end{equation}

A crossover occurs when

\begin{equation}
E_{\downarrow\downarrow}\sim E_{\rm AFM}.
\end{equation}

Beyond this regime, the channel ground state becomes

\begin{equation}
|\chi_0\rangle=
a|\downarrow\downarrow\rangle
+b|\uparrow\downarrow\rangle
+c|\downarrow\uparrow\rangle
+d|\uparrow\uparrow\rangle,
\label{eq:channel_general}
\end{equation}

where the coefficients depend on both $B$ and $\theta_{\rm so}$.

For sufficiently large fields satisfying

\begin{equation}
|h_z|\gg J,
\qquad
|h_x|,|h_y|\ll |h_z|,
\end{equation}

the transverse mixing becomes perturbatively small and

\begin{equation}
|\chi_0\rangle\rightarrow |\downarrow\downarrow\rangle.
\end{equation}

Thus the channel evolves from an antiferromagnetic state at low fields to a spin-polarized state at large fields.

\subsection{Loss of ground-state overlap in the full chain}

As discussed before, $B=0$, the full four-spin Hamiltonian preserves time-reversal symmetry,

\begin{equation}
\mathcal TH\mathcal T^{-1}=H,
\end{equation}

as well as reflection symmetry

\begin{equation}
1\leftrightarrow4,
\qquad
2\leftrightarrow3.
\end{equation}

The initial state therefore overlaps with multiple low-energy eigenstates,

\begin{equation}
|\psi_0\rangle=\sum_n c_n|E_n\rangle.
\end{equation}

Introducing the magnetic field breaks time-reversal symmetry through

\begin{equation}
\sum_i(g_i\mathbf B)\cdot\mathbf S_i,
\end{equation}

and drives the full-chain ground state toward

\begin{equation}
|\downarrow\downarrow\downarrow\downarrow\rangle.
\end{equation}

Consequently,

\begin{equation}
|\langle E_0|\psi_0\rangle|^2\rightarrow0,
\end{equation}

and the initial state projects predominantly onto the first two excited states, $\ket{E_1}=\ket{\uparrow\downarrow\downarrow\downarrow}$ and $\ket{E_2}=\ket{\downarrow\downarrow\downarrow\uparrow}$

\begin{equation}
|\psi_0\rangle
\approx
c_1|E_1\rangle
+
c_2|E_2\rangle.
\label{eq:two_state_projection}
\end{equation}

The transport problem therefore reduces to an effective two-level system.

\subsection{Emergence of the transport doublet}

Let $|\chi_0\rangle_{23}$ denote the channel ground state. We define

\begin{equation}
|L\rangle=
|\uparrow\rangle_1
\otimes
|\chi_0\rangle_{23}
\otimes
|\downarrow\rangle_4,
\end{equation}

\begin{equation}
|R\rangle=
|\downarrow\rangle_1
\otimes
|\chi_0\rangle_{23}
\otimes
|\uparrow\rangle_4.
\end{equation}

Projecting onto this subspace gives

\begin{equation}
H_{\rm eff}
=
\begin{pmatrix}
\langle L|H|L\rangle & \langle L|H|R\rangle
\\
\langle R|H|L\rangle & \langle R|H|R\rangle
\end{pmatrix}.
\end{equation}

The detuning is

\begin{equation}
\Delta=
\langle L|H|L\rangle-
\langle R|H|R\rangle,
\label{eq:detuning}
\end{equation}

while

\begin{equation}
t_{\rm eff}
=
\langle L|H|R\rangle
\end{equation}

defines the effective tunneling amplitude.

High-fidelity transfer requires

\begin{equation}
|\Delta|\ll |t_{\rm eff}|,
\end{equation}

for which

\begin{equation}
|E_{\pm}\rangle
\approx
\frac{|L\rangle\pm|R\rangle}{\sqrt{2}}.
\end{equation}

The transfer time is

\begin{equation}
t^*=
\frac{\pi}{E_+-E_-}.
\end{equation}

\subsection{Resonance at $\theta_{\rm so}=\pi$}

We now explicitly evaluate the detuning $\Delta$ defined in Eq.~\eqref{eq:detuning}  now for $\theta_{\rm so}=\pi$.

The left bond interaction is

\begin{equation}
H_{12}
=
j\sum_{a,b=x,y,z}
S_1^aR_{ab}S_2^b.
\end{equation}

For the state $|L\rangle$, site $1$ is fixed in the state $|\uparrow\rangle$, implying

\begin{equation}
\langle S_1^x\rangle=
\langle S_1^y\rangle=0,
\end{equation}

\begin{equation}
\langle S_1^z\rangle=\frac{1}{2}.
\end{equation}

Therefore,

\begin{equation}
\langle L|H_{12}|L\rangle
=
\frac{j}{2}
\sum_{b=x,y,z}
R_{zb}\langle S_2^b\rangle.
\label{eq:leftbond_explicit}
\end{equation}

Similarly, the right bond interaction is

\begin{equation}
H_{34}
=
j\sum_{a,b=x,y,z}
S_3^aR_{ab}S_4^b.
\end{equation}

For the state $|L\rangle$, site $4$ is fixed in $|\downarrow\rangle$, giving

\begin{equation}
\langle S_4^x\rangle=
\langle S_4^y\rangle=0,
\end{equation}

\begin{equation}
\langle S_4^z\rangle=-\frac{1}{2}.
\end{equation}

Thus,

\begin{equation}
\langle L|H_{34}|L\rangle
=
-\frac{j}{2}
\sum_{a=x,y,z}
R_{az}\langle S_3^a\rangle.
\label{eq:rightbond_explicit}
\end{equation}

For the state $|R\rangle$, the spin orientations on sites $1$ and $4$ are reversed, causing the signs of both contributions to change.

Subtracting the two diagonal matrix elements therefore yields

\begin{equation}
\Delta=
j
\sum_{b=x,y,z}
R_{zb}\langle S_2^b\rangle
-
j
\sum_{a=x,y,z}
R_{az}\langle S_3^a\rangle.
\label{eq:delta_general}
\end{equation}

At
$\theta_{\rm so}=\pi$,

the rotation matrix simplifies to

\begin{equation}
R(\hat n,\pi)=2\hat n\hat n^T-I,
\end{equation}

which is manifestly symmetric, $R_{ab}=R_{ba}$.

The two-spin channel Hamiltonian also remains symmetric under site exchange $
2\leftrightarrow3,$
which implies

\begin{equation}
\langle S_2^\alpha\rangle=
\langle S_3^\alpha\rangle.
\end{equation}

Substituting these relations into Eq.~(\ref{eq:delta_general}) gives

\begin{equation}
\boxed{
\Delta=0
}
\end{equation}

exactly.

The two transport states therefore become exactly resonant, and the remaining splitting of the first two excited states is controlled solely by the off-diagonal tunneling matrix element $t_{\rm eff}$. 
This explains the robust emergence of an isolated near-degenerate doublet and the resulting high-fidelity state transfer observed numerically at $\theta_{\rm so}=\pi$. 

Finally, we note that $\theta_{\rm so}=\pi$ is not the only parameter regime in which enhanced transport can occur. More generally, high-fidelity transfer requires the effective detuning $\Delta$ to remain small compared to the effective tunneling matrix element $t_{\rm eff}$. 
Therefore, for other choices of spin-orbit angle, rotation axis, or magnetic field strength, near-resonances can also occur when $|\Delta|\ll |t_{\rm eff}|$,
leading to enhanced transport even away from $\theta_{\rm so}=\pi$. This explains the additional high-fidelity regions observed numerically in Fig.~\ref{figmag_new}.

The particularly robust regime emerges when the spin-orbit axis is aligned with the magnetic field direction,

\begin{equation}
\hat n=\hat z,
\end{equation}

with the external field also applied along the laboratory $z$ axis. In this case,

\begin{equation}
R(\hat z,\theta_{\rm so})=
\begin{pmatrix}
\cos\theta_{\rm so} & -\sin\theta_{\rm so} & 0
\\
\sin\theta_{\rm so} & \cos\theta_{\rm so} & 0
\\
0 & 0 & 1
\end{pmatrix},
\end{equation}

which immediately implies

\begin{equation}
R_{xz}=R_{yz}=R_{zx}=R_{zy}=0.
\end{equation}

As a result, all exchange-induced single-spin flip processes vanish identically.

Furthermore, for experimentally relevant $g$ tensors where

\begin{equation}
g_{zz}\gg g_{xz},g_{yz},
\end{equation}

the effective transverse Zeeman fields satisfy

\begin{equation}
h_x\approx h_y\approx0.
\end{equation}

The channel Hamiltonian therefore becomes approximately block diagonal, strongly suppressing hybridization between antiferromagnetic and ferromagnetic sectors. Consequently, the transport doublet remains well isolated over a broader range of $\theta_{\rm so}$, leading to the enhanced robustness observed numerically for spin-orbit axes aligned with the magnetic field.

\begin{figure*}[t]
    \centering
    \includegraphics[width=0.98\linewidth]{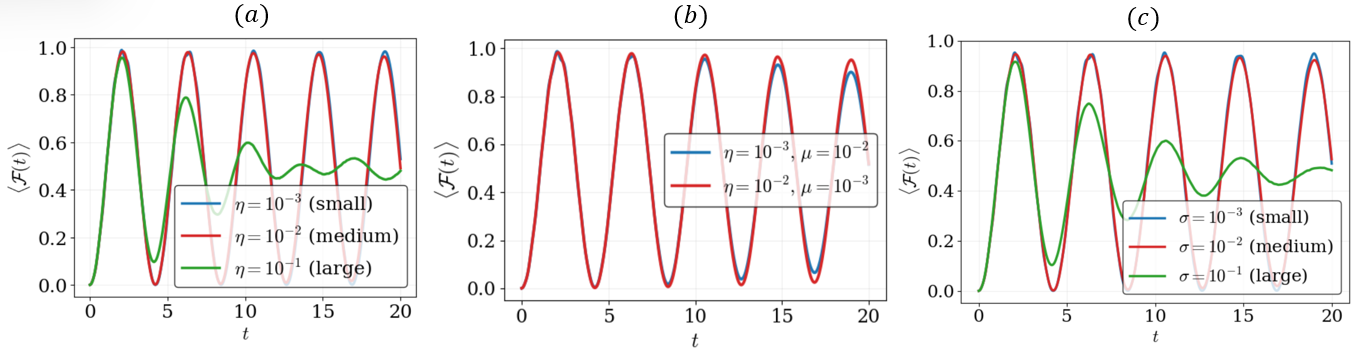}
\caption{
Disorder-averaged state-transfer fidelity in the presence of quasi-static charge noise for three representative noise models. 
(a) Correlated (uniform) noise applied equally to all exchange couplings, shown for $\delta J/J \sim 10^{-3}, 10^{-2}, 10^{-1}$; increasing noise broadens and suppresses fidelity peaks due to run-to-run fluctuations in the oscillation frequency. 
(b) Independent noise on the intra-channel exchange $J$ and the end-channel coupling $j$ ($\eta \neq \mu$); fluctuations in $j$ degrade fidelity more strongly than comparable fluctuations in $J$, reflecting the sensitivity of the optimal transfer time to variations in the end-channel coupling. 
(c) Fully uncorrelated noise, where each coupling fluctuates independently ($\eta \rightarrow \eta_i$, $\mu \rightarrow \mu_i$); within the relevant time window, the fidelity remains qualitatively similar to the correlated case, indicating that spatial inhomogeneities do not significantly degrade state transfer.
}
   \label{fig10}
\end{figure*}

\section{Effective Hamiltonian with $B=0$ from Van Vleck perturbation theory}
\label{app3}

To obtain the effective interaction between the end qubits in the absence of magnetic field, we employ Van Vleck perturbation theory to eliminate the high-energy sector and derive an effective Hamiltonian within the low-energy subspace relevant for state transfer.

We decompose the Hamiltonian as
\begin{equation}
H = H_0 + V,
\end{equation}
where $H_0$ is diagonal in the basis of singlet and triplet states of the channel, and $V$ describes the coupling between the end qubits and the channel.

Since the dynamics occurs within the single-excitation sector ($S_z=0$), we consider the basis
\begin{equation}
\{|\uparrow S \downarrow\rangle,\;
|\downarrow S \uparrow\rangle,\;
|\uparrow T_0 \downarrow\rangle,\;
|\downarrow T_0 \uparrow\rangle,\;
|\uparrow T_- \uparrow\rangle,\;
|\downarrow T_+ \downarrow\rangle\}.
\end{equation}

The low-energy subspace is spanned by 
$\{|\uparrow S \downarrow\rangle, |\downarrow S \uparrow\rangle\}$, while the remaining states form the high-energy sector.

Following the Van Vleck formalism, the block-diagonal effective Hamiltonian is obtained as
\begin{equation}
H_{\mathrm{eff}} = P H P + P V \frac{1}{E - Q H_0 Q} V P + \mathcal{O}(V^3),
\end{equation}
where $P$ projects onto the low-energy subspace and $Q=1-P$ onto the complementary sector.

More explicitly, using the standard Van Vleck expansion, the matrix elements of the effective Hamiltonian up to second order are
\begin{equation}
(H_{\mathrm{eff}})_{mn}
=
H_{mn}
+
\sum_{k \in Q}
\frac{H_{mk} H_{kn}}{E_m - E_k}
+
\sum_{k \in Q}
\frac{H_{mk} H_{kn}}{E_n - E_k},
\end{equation}
where $m,n$ belong to the low-energy subspace and $k$ runs over intermediate (high-energy) states.

Applying this to our system, the diagonal terms ($G_{11}, G_{22}$) receive second-order corrections, while the off-diagonal terms ($G_{12}, G_{21}$) determine the effective transition amplitude between 
$|\uparrow S \downarrow\rangle$ and $|\downarrow S \uparrow\rangle$.

Evaluating these contributions explicitly (as implemented in the symbolic calculation shown in the text), we obtain the effective Hamiltonian in the low-energy subspace as
\begin{equation}
H_{\mathrm{eff}} =
\begin{pmatrix}
-J_{xx}-J_{yy}-J_{zz} + \frac{G_{11}}{2} & T \\
T^* & -J_{xx}-J_{yy}-J_{zz} + \frac{G_{22}}{2}
\end{pmatrix},
\end{equation}
where the off-diagonal matrix element $T$ arises from second- and higher-order virtual processes through the triplet sector.

The dominant contribution to the transition amplitude is given by
\begin{equation}
T \sim
\frac{(j_{xx}+j_{yy})^2 \big[2 i J_{zz} + J_{xx} + J_{yy} + i(J_{xy}-J_{yx})\big]}
{(J_{xx}+J_{yy}) \big[-2 i J_{zz} + J_{xx} + J_{yy} + 2 J_{zz}\big]},
\end{equation}
which sets the frequency of coherent oscillations between the two end-spin configurations. And the diagonal term gets modified by,
\begin{eqnarray}
G_{11} = G_{22} =&
- J_{xx} - J_{yy} - J_{zz}
- \frac{2 j_{zz}^2 (J_{xy} - J_{yx})^2}{J_{xx} + J_{yy}}\\
-& \frac{(j_{xy} - j_{yx})^2 (j_{xx} + j_{yy})^2}
{-2 j_{zz} + J_{xx} + J_{yy} + 2 J_{zz}} .
\end{eqnarray}

This expression results from summing over all virtual transitions through the triplet manifold, with the denominators encoding the energy cost of accessing intermediate states and the numerators capturing the structure of anisotropic exchange couplings. The Van Vleck transformation thus provides a controlled perturbative framework to derive the effective two-level dynamics governing state transfer.

\section{Robustness against noise}
\label{app4}

Charge noise is the dominant low-temperature decoherence mechanism in gate-defined semiconductor spin qubits, arising from fluctuations in the electrostatic environment such as charge traps, defects, and two-level fluctuators \cite{noise1}. Since exchange interactions are electrically controlled, these fluctuations directly translate into noise in the coupling strengths.

To leading order, voltage fluctuations induce relative variations in the exchange couplings,
\begin{equation}
J(V) \approx J_0 + \frac{\partial J}{\partial V}\,\delta V,
\qquad
J \rightarrow (1+\eta_i)J,
\end{equation}
where $\eta_i = \delta J_i/J_i$ is a dimensionless random variable. In the quasi-static approximation, appropriate for low-frequency $1/f$ noise, these parameters remain fixed during a single realization of time evolution but vary between experimental runs. Observables are therefore obtained via disorder averaging.

In our architecture, all exchange couplings are electrically controlled and thus subject to fluctuations,
\begin{equation}
J_{\alpha\beta}^{(i)} \rightarrow (1+\eta_i)\, J_{\alpha\beta}^{(i)},
\qquad
j_{\alpha\beta}^{(i)} \rightarrow (1+\mu_i)\, j_{\alpha\beta}^{(i)},
\end{equation}
where $i$ labels the corresponding bond.

We analyze three representative noise models, shown in Fig.~\ref{fig10}, for noise strengths $\delta J/J \sim 10^{-3}, 10^{-2}, 10^{-1}$:

\textit{(i) Correlated (uniform) noise.}  
All couplings fluctuate identically, $\eta_i = \mu_i = \eta$, corresponding to a global rescaling of the energy scale. As shown in Fig.~\ref{fig10}(a), increasing noise suppresses and broadens fidelity oscillations. This originates from run-to-run fluctuations in the effective oscillation frequency, which shift the optimal transfer time and lead to dephasing upon averaging.

\textit{(ii) Independent noise on intra- and end-channel couplings.}  
We next allow distinct fluctuations for intra-channel exchange $J$ and end-channel coupling $j$, i.e., $\eta \neq \mu$. As shown in Fig.~\ref{fig10}(b), fluctuations in $j$ degrade the fidelity more strongly than comparable fluctuations in $J$, reflecting the sensitivity of the optimal transfer time $t_{\mathrm{opt}} \sim 1/j$ to variations in the end-channel coupling.

\textit{(iii) Fully uncorrelated noise.}  
Finally, we consider the most general case in which each coupling fluctuates independently across the array, $\eta_i$ and $\mu_i$. As shown in Fig.~\ref{fig10}(c), within the experimentally relevant time window ($t \leq 10$), the disorder-averaged fidelity remains qualitatively similar to the correlated-noise case. In particular, the overall oscillatory structure and achievable peak fidelity are largely preserved.

Across all noise models, the dominant effect of charge noise is governed by fluctuations in the overall energy scale, while spatial inhomogeneities do not significantly further degrade state transfer on experimentally relevant timescales. Consequently, high-fidelity quantum communication remains achievable for realistic noise levels $\delta J/J \sim 10^{-3} - 10^{-1}$.

\bibliography{references}
\end{document}